\documentstyle[12pt]{article}
\def\beq{\begin{equation}}
\def\eeq{\end{equation}}
\def\bea{\begin{eqnarray}}
\def\eea{\end{eqnarray}}
\def\bq{\begin{quote}}    
\def\eq{\end{quote}}

\def\bq{\begin{quote}}
\def\eq{\end{quote}}

\def\bq{\begin{quote}}
\def\eq{\end{quote}}

\def \lsim{\mathrel{\vcenter
     {\hbox{$<$}\nointerlineskip\hbox{$\sim$}}}}
\def \gsim{\mathrel{\vcenter
     {\hbox{$>$}\nointerlineskip\hbox{$\sim$}}}}

\def\gappeq{\mathrel{\rlap {\raise.5ex\hbox{$>$}}
{\lower.5ex\hbox{$\sim$}}}}

\def\lappeq{\mathrel{\rlap{\raise.5ex\hbox{$<$}}
{\lower.5ex\hbox{$\sim$}}}}

\def\bbz{fa Z \kern-8.9pt Z}

\begin{document}

\baselineskip 24pt
\newcommand{\sheptitle}
{Large or Small Angle MSW from Single Right-Handed Neutrino Dominance}

\newcommand{\shepauthor}
{S. F. King}

\newcommand{\shepaddress}
{Department of Physics and Astronomy,
University of Southampton, Southampton, SO17 1BJ, U.K.}

\newcommand{\shepabstract}
{In this talk we discuss a natural explanation
of both neutrino mass hierarchies 
{\it and} large neutrino mixing angles,
as required by the atmospheric neutrino data, in terms of
a single right-handed neutrino giving the
dominant contribution to the 23 block of the light effective
neutrino matrix, and illustrate this mechanism
in the framework of models with $U(1)$ family symmetries.
Sub-dominant contributions from other right-handed neutrinos are required to
give small mass splittings appropriate to the MSW
solution to the solar neutrino problem. 
We present three explicit examples for achieving 
the small angle MSW solution in the framework of
$U(1)$ family symmetry models
containing three right-handed neutrinos, 
which can naturally describe all quark and lepton masses
and mixing angles. In this talk we also extend the analysis
to the large angle MSW solution.}

\begin{titlepage}
\begin{flushright}
hep-ph/9908476\\
\end{flushright}
\begin{center}
{\large{\bf \sheptitle}}
\footnote{To appear in the proceedings of the 6th San Miniato
Topical Seminar on ``Neutrino and Astroparticle Physics'',
Nuclear Physics B (Proc. Suppl.)}
\bigskip \\ \shepauthor \\ \mbox{} \\ {\it \shepaddress} \\ \vspace{.5in}
{\bf Abstract} \bigskip \end{center} \setcounter{page}{0}
\shepabstract
\end{titlepage}

There is now strong evidence for atmospheric neutrino oscillations
\cite{SK}.
The most recent analyses of Super-Kamiokande \cite{SK} involve
the hypothesis of
$\nu_{\mu}\rightarrow \nu_{\tau}$ oscillations
with maximal mixing $\sin^2 2\theta_{23} =1$
and a mass splitting of
$\Delta m_{23}^2 = 2.2\times 10^{-3}\ eV^2$.
Using all their data sets
analysed in different ways they quote $\sin^2 2\theta_{23} > 0.82$
and a mass splitting of
$1.5\times 10^{-3}\ eV^2 <\Delta m_{23}^2 < 6\times 10^{-3}\ eV^2$
at 90\% confidence level. 

The evidence for solar neutrino oscillations is almost as strong.  
There are
a panoply of experiments looking at different energy ranges, and
the best fit to all of them has been narrowed down to two basic
scenarios corresponding to either resonant oscillations
$\nu_e \rightarrow \nu_0$ (where for example 
$\nu_0$ may be a linear combination
of $\nu_{\mu} , \nu_{\tau}$) inside the Sun
(MSW \cite{MSW}) or ``just-so'' oscillations in the vacuum between the Sun
and the Earth \cite{justso1}, \cite{justso2}.
There are three MSW fits and one vacuum oscillation fit:

(i) the small angle MSW 
solution is $\sin^2 2\theta_{12} \approx 5 \times 10^{-3}$
and $\Delta m_{12}^2 \approx  5\times 10^{-6}\ eV^2$;

(ii) the large angle MSW solution is $\sin^2 2\theta_{12} \gsim 0.2$
and $\Delta m_{12}^2 \approx  1.8\times 10^{-5}\ eV^2$;

(iii) an additional MSW large angle solution exists with a lower probability 
\cite{BKS};

(iv) The vacuum oscillation solution is $\sin^2 2\theta_{12} \approx 0.75$
and $\Delta m_{12}^2 \approx  6.5\times 10^{-11}\ eV^2$ \cite{BKS}.

The standard model has zero neutrino masses, so any indication of
neutrino mass is very exciting since it represents
new physics beyond the standard model. In this paper
we shall assume the see-saw mechanism and no light sterile neutrinos.
The see-saw mechanism \cite{seesaw} implies that the 
three light neutrino masses
arise from some heavy ``right-handed neutrinos'' $N^p_R$ 
(in general there can be $Z$ gauge singlets with $p=1,\ldots Z$)
with a $Z\times Z$ Majorana mass matrix 
$M^{pq}_{RR}$ whose entries take values
at or below the unification scale $M_U \sim 10^{16}$ GeV.
The presence of electroweak scale Dirac mass terms $m_{LR}^{ip}$ 
(a $3 \times Z$ matrix) connecting the
left-handed neutrinos $\nu^i_L$ ($i=1,\ldots 3$)
to the right-handed neutrinos $N^p_R$ 
then results in a very light see-saw suppressed effective $3\times 3$ Majorana
mass matrix 
\beq
m_{LL}=m_{LR}M_{RR}^{-1}m_{LR}^T
\label{seesaw}
\eeq
for the left-handed neutrinos $\nu_L^i$, which are the light physical
degrees of freedom observed by experiment. 

Not surprisingly, following
the recent data,
there has been a torrent of theoretical papers concerned with
understanding how to extend the standard model in order to accomodate
the atmospheric and solar neutrino data. 
Perhaps the minimal extension of the standard model capable of accounting
for the atmospheric neutrino data involves
the addition of a {\em single} right-handed neutrino $N_R$ \cite{SK1},
\cite{SK2}.
This is a special case 
of the general see-saw model with $Z=1$, so that
$M_{RR}$ is a trivial $1 \times 1$ matrix
and $m_{LR}$ is a $3 \times 1$ column matrix where
$m_{LR}^T=(\lambda_{\nu_e}, \lambda_{\nu_{\mu}},
\lambda_{\nu_{\tau}})v_2$ with
$v_2$ the vacuum expectation value of the Higgs field $H_2$
which is responsible for the neutrino Dirac masses,
and the notation for the Yukawa couplings $\lambda_i$
indicates that we are in
the charged lepton mass eigenstate basis $e_L, \mu_L, \tau_L$
with corresponding neutrinos $\nu_{e_L}, \nu_{\mu_L}, \nu_{\tau_L}$.
Since $M_{RR}$ is trivially invertible 
the light effective mass matrix in Eq.\ref{seesaw}
in the $\nu_{e_L}, \nu_{\mu_L}, \nu_{\tau_L}$ basis is simply given by
\beq
m_{LL} =
\left( \begin{array}{lll}
 \lambda_{\nu_e}^2 &  \lambda_{\nu_e} \lambda_{\nu_{\mu}}
 & \lambda_{\nu_e} \lambda_{\nu_{\tau}}     \\
\lambda_{\nu_e} \lambda_{\nu_{\mu}} & \lambda_{\nu_{\mu}}^2 
& \lambda_{\nu_{\mu}} \lambda_{\nu_{\tau}} \\
\lambda_{\nu_e }\lambda_{\nu_{\tau}} 
& \lambda_{\nu_{\mu}} \lambda_{\nu_{\tau}} & \lambda_{\nu_{\tau}}^2
\end{array}
\right)\frac{v_2^2}{M_{RR}}.
\label{matrix}
\eeq
The matrix in Eq.\ref{matrix} has vanishing determinant which implies 
a zero eigenvalue. Furthermore the submatrix in the 23
sector has zero determinant which implies a second zero eigenvalue
associated with this sector.
In order to account for the Super-Kamiokande data we assumed \cite{SK1}:
\beq
\lambda_{\nu_e} \ll \lambda_{\nu_{\mu}} \approx \lambda_{\nu_{\tau}}.
\label{hierarchy}
\eeq
In the $\lambda_{\nu_e}=0$ limit the matrix in Eq.\ref{matrix} has zeros
along the first row and column, and so clearly
$\nu_e$ is massless, and the other two
eigenvectors are simply
\beq
\left(
\begin{array}{l}
\nu_0 \\
\nu_3
\end{array}
\right)
=
\left(
\begin{array}{ll}
c_{23} & -s_{23}\\
s_{23} & c_{23}
\end{array}
\right)
\left(
\begin{array}{l}
\nu_{\mu} \\
\nu_{\tau}
\end{array}
\right)
\label{bbasis}
\eeq
where $t_{23}=\lambda_{\nu_{\mu}}/\lambda_{\nu_{\tau}}$, with
$\nu_0$ being massless, due to the vanishing
of the determinant of the 23 submatrix and $\nu_3$ having a
mass $m_{\nu_3}=(\lambda_{\nu_{\mu}}^2 + \lambda_{\nu_{\tau}}^2)
v_2^2/M_{RR}$.
The Super-Kamiokande data is accounted for by choosing the parameters
such that
$t_{23} \sim 1$ and $m_{\nu_3} \sim 5\times 10^{-2}$ eV.
In this approximation 
the atmospheric neutrino data is then consistent with
$\nu_{\mu}\rightarrow \nu_{\tau}$ oscillations via two state mixing,
between $\nu_3$ and $\nu_0$. Note how the single right-handed neutrino
coupling to the 23 sector implies vanishing determinant of the 23
submatrix. This provides a natural explanation of
both large 23 mixing angles and a hierarchy
of neutrino masses in the 23 sector at the same time \cite{SK1}.

In order to account for the solar neutrino data a small mass perturbation
is required to lift the massless degeneracy
of the two neutrinos $\nu_0 ,\nu_e$. 
In our original approach \cite{SK1}
\footnote{
Another approach \cite{SK2}
which does not rely on additional right-handed neutrinos
is to use SUSY radiative corrections so that the one-loop corrected
neutrino masses are not zero but of order $10^{-5}$ eV suitable
for the vacuum oscillation solution.} we
introduced additional right-handed neutrinos in order to provide
a subdominant contribution to the effective mass matrix
in Eq.\ref{matrix}. To be precise we assumed a single dominant
right-handed neutrino below the unification scale, with additional
right-handed neutrinos at the unification scale which lead to 
subdominant contributions to the effective neutrino mass matrix.
By appealing to quark and lepton mass hierarchy we assumed
that the additional subdominant right-handed neutrinos generate 
a contribution 
$m_{\nu_{\tau}}\approx m_t^2/M_U \approx 2\times 10^{-3}$ eV,
where $m_t$ is the top quark mass.
The effect of this is to give a mass perturbation
to the 33 component of the mass matrix in Eq.\ref{matrix},
which results in $\nu_0 $ picking up a small mass, 
through its $\nu_{\tau}$ component, while $\nu_e $ remains
massless. Solar neutrino oscillations then arise from
$\nu_e \rightarrow \nu_0$ with the mass
splitting in the right range
for the small angle MSW solution, controlled by a small mixing angle
$\theta_{12} \approx 
\lambda_{\nu_e}/\sqrt{\lambda_{\nu_{\mu}}^2 + \lambda_{\nu_{\tau}}^2}$.
The main prediction of this scheme is of the neutrino oscillation
$\nu_e \rightarrow \nu_3$ with a mass difference 
$\Delta m_{13}^2 \approx \Delta m_{23}^2$
determined by the
Super-Kamiokande data and a
mixing angle $\theta_{13} \approx \theta_{12}$
determined by the small angle
MSW solution. Such oscillations may be observable at the proposed
long baseline experiments via $\nu_3 \rightarrow \nu_e$
which implies 
$\nu_{\mu} \rightarrow \nu_e$ oscillations
with $\sin^2 2\theta \approx 5 \times 10^{-3}$ (the small MSW angle)
and $\Delta m^2 \approx  2.2\times 10^{-3}\ eV^2$ (the
Super-Kamiokande square mass difference).

It should be clear from the foregoing discussion that the motivation for
single right-handed neutrino dominance (SRHND) is that
the determinant of the 23 submatrix of Eq.\ref{matrix}
approximately vanishes, leading to a natural explanation of {\em both} large
neutrino mixing angles {\em and} hierarchical neutrino masses in the 23
sector {\em at the same time} \cite{SK1}. Although the explicit example of
SRHND above was based on one of the right-handed neutrinos being lighter
than the others, it is clear that the idea of SRHND is more general
than this. 

In \cite{SK3} we defined SRHND more generally
as the requirement that a single right-handed neutrino gives the
dominant contribution to the 23 submatrix of the light effective
neutrino mass matrix (which can be achieved in other ways
than one of the right-handed neutrinos being lighter than the others.)
We addressed
the following two questions:
\newline
1. What are the general conditions under
which SRHND in the 23 block can arise and 
how can we quantify the contribution of the sub-dominant 
right-handed neutrinos which are responsible
for breaking the massless degeneracy, and allowing the small angle MSW
solution? \newline
2. How can we understand the pattern of neutrino Yukawa couplings in
Eq.\ref{hierarchy} where the assumed equality 
$\lambda_{\nu_{\mu}} \approx \lambda_{\nu_{\tau}}$
is apparently at odds with the hierarchical Yukawa
couplings in the quark and charged lepton sector? \newline

In order to address the two questions above we discuss
SRHND in the context of a $U(1)$ family symmetry. In ref.\cite{SK3}
we gave general conditions that theories with $U(1)$ family
symmetry must satisfy in order to have SRHND and showed that 
the models in \cite{Altarelli} satisfy these conditions.
In this talk we briefly review this approach, giving three
examples based on the general analysis in ref.\cite{SK3}.

The Wolfenstein parametrisation of the CKM matrix is roughly
\beq
V_{CKM} \sim
\left( \begin{array}{lll}
1 &  \lambda & \lambda^3 \\
\lambda & 1 & \lambda^2 \\
\lambda^3 & \lambda^2 & 1
\end{array}
\right)
\eeq
With a single dominant right-handed neutrino we expect 
equal neutrino mixing angles in 12 and 13 sectors
\beq
\theta_{12} \sim \theta_{13}
\eeq
CHOOZ \cite{CHOOZ} 
tells us that over most of the interesting mass range
$\sin^2 2 \theta_{13}<0.18$, corresponding to
$\theta_{13}\leq \lambda$.
Thus there are two interesting 
possibilities for the choice of angle, corresponding to
large or small angle MSW with 
$\theta^{large}_{13}\sim \theta^{large}_{12} \sim \lambda$, or 
$\theta^{small}_{13} \sim \theta^{small}_{12} \sim \lambda^2$, with
the Maki-Nakagawa-Sakata matrix, 
the leptonic analogue of the CKM matrix,
determined in each case:
\beq
V^{large}_{MNS} \sim
\left( \begin{array}{lll}
1 &  \lambda & \lambda \\
\lambda & 1 & 1 \\
\lambda & 1 & 1
\end{array}
\right),
V^{small}_{MNS} \sim
\left( \begin{array}{lll}
1 &  \lambda^2 & \lambda^2 \\
\lambda^2 & 1 & 1 \\
\lambda^2 & 1 & 1
\end{array}
\right)
\eeq
Note that in the large angle MSW case, we are relying on factors 
of order unity implicitly present in each element to give us a large enough
MSW angle without violating the CHOOZ constraint.
Our working assumption is that $V^{large}_{MNS}$ or $V^{small}_{MNS}$
originates from both the
neutrino sector and the charged lepton sector in roughly equal
measure which, together with Eq.\ref{matrix}, gives 
\beq
m^{large}_{LL} \sim
\left( \begin{array}{lll}
\lambda^2 &  \lambda & \lambda\\
\lambda & 1 & 1 \\
\lambda & 1 & 1
\end{array}
\right)m_{\nu_3},
m^{small}_{LL} \sim
\left( \begin{array}{lll}
\lambda^4 &  \lambda^2 & \lambda^2\\
\lambda^2 & 1 & 1 \\
\lambda^2 & 1 & 1
\end{array}
\right)m_{\nu_3}
\eeq
corresponding to $V^{large}_{MNS}$, $V^{small}_{MNS}$, respectively.
In general would expect $m_{\nu_2} \sim m_{\nu_3}$ due to
large 23 mixing, but with a single right-handed neutrino
the vanishing determinant of 23 block solves
this problem by setting  $m_{\nu_2}=0$ .
In order obtain the desired hierarchy between the 
MSW neutrino mass and the atmospheric neutrino mass,
$m_{\nu_2} /m_{\nu_3} \sim \lambda^2$, we must add extra subdominant
right-handed neutrinos which contribute to the 23 block at order 
$O(\lambda^2)$. In the small angle case this would lead to 
$m_{\nu_1}/m_{\nu_2}\sim \lambda^2$ and a hierarchy of neutrino
masses, while in the large angle case we
would have $m_{\nu_1} \lsim m_{\nu_2}$, leading to a semi-hierarchical
neutrino mass pattern.

To proceed we introduce a $U(1)$ family
symmetry of the kind suggested by Ibanez and Ross \cite{IR}.
For example a suitable choice of quark, lepton and Higgs charges 
leads to the quark and charged lepton Yukawa matrices \cite{Ramond}: 
\beq
Y^u \sim
\left( \begin{array}{lll}
\lambda^8 &  \lambda^5 & \lambda^3 \\
\lambda^7 &  \lambda^4 & \lambda^2 \\
\lambda^5 &  \lambda^2 & 1
\end{array}
\right), 
Y^d \sim
\left( \begin{array}{lll}
\lambda^4 &  \lambda^3 & \lambda^3 \\
\lambda^3 & \lambda^2 & \lambda^2 \\
\lambda & 1 & 1
\end{array}
\right)\lambda^n 
\eeq
\beq
Y_{large}^e \sim
\left( \begin{array}{lll}
\lambda^4 & \lambda^3 &  \lambda  \\
\lambda^3 & \lambda^2 & 1 \\
\lambda^3 & \lambda^2 & 1
\end{array}
\right) \lambda^n,
Y_{small}^e \sim
\left( \begin{array}{lll}
\lambda^4 &  \lambda^4 &  \lambda^2  \\
\lambda^2 & \lambda^2 & 1 \\
\lambda^2 & \lambda^2 & 1
\end{array}
\right) \lambda^n
\eeq
which lead to the following successful quark and lepton
mass relations:
\beq
\frac{m_u}{m_t} \sim \lambda^8, \ \ 
\frac{m_c}{m_t} \sim \lambda^4, \ \ 
\frac{m_d}{m_b} \sim \lambda^4, \ \ 
\frac{m_s}{m_b} \sim \lambda^2, 
\eeq
\beq
\frac{m_e}{m_{\tau}} \sim \lambda^4, \ \ 
\frac{m_{\mu}}{m_{\tau}} \sim \lambda^2. \ \ 
\eeq
\beq
\frac{m_b}{m_t} \sim \lambda^3, \ \ 
\frac{m_b}{m_{\tau}} \sim 1, \ \ 
\eeq
where the last relations are valid in the MSSM at the unification
scale, where there are two Higgs doublets with vacuum expectation
values $v_1,v_2$ coupling to down-type quarks, up-type quarks,
respectively, and $\tan \beta = v_2/v_1 \sim \lambda^{n-3}$.
The correct CKM matrix given earlier is also reproduced,
and $V^{large}_{MNS}$, $V^{small}_{MNS}$ are consistent with 
$Y_{large}^e$, $Y_{small}^e$, respectively.

When dealing with the lepton charges, it is convenient to
absorb the physical Higgs charge $h_u$ into the physical lepton
charges $l_i$, whereupon we find the redefined lepton charges \cite{SK3} 
\beq
 \left( \begin{array}{c} \nu_{1L}\\ e_{1L} \end{array} \right),
 \left( \begin{array}{c} \nu_{2L}\\ e_{2L} \end{array} \right),
 \left( \begin{array}{c} \nu_{3L}\\ e_{3L} \end{array} \right)        
=(m+l_3,l_3,l_3)
\label{li}
\eeq
where $m=1,2$ for $Y_{large}^e$, $Y_{small}^e$ cases, respectively,
and the numerical value of $l_3$ remains a free choice,
which is specified precisely in the examples below.

We now give three examples of $U(1)$ charge assignments
for the three lepton doublets and three right-handed neutrinos
which satisfies SRHND for the small angle MSW case $m=2$ \cite{SK3}.
We classify the cases according to the upper block
structure of the resulting heavy Majorana matrix:

(i) ``Diagonal dominated'' upper block of heavy Majorana matrix.
\beq
 \left( \begin{array}{c} \nu_{1L}\\ e_{1L} \end{array} \right),
 \left( \begin{array}{c} \nu_{2L}\\ e_{2L} \end{array} \right),
 \left( \begin{array}{c} \nu_{3L}\\ e_{3L} \end{array} \right)
=(1/2,-3/2,-3/2)
\eeq
\beq
\bar{\nu}_{1R},\bar{\nu}_{2R},\bar{\nu}_{3R} = (0,1,2)
\eeq

The resulting mass matrices are:

\beq
M_{RR} \sim
\left( \begin{array}{lll}
1 &  \lambda & \lambda^2\\
\lambda & \lambda^2 & \lambda^3 \\
\lambda^2 & \lambda^3 & \lambda^4
\end{array}
\right)M
\eeq

\beq
m_{LR} \sim
\left( \begin{array}{lll}
\lambda^{1/2} & \lambda^{3/2} & \lambda^{5/2}\\
\lambda^{3/2} & \lambda^{1/2} & \lambda^{1/2} \\
\lambda^{3/2} & \lambda^{1/2} & \lambda^{1/2}
\end{array}
\right)v_2
\eeq

\beq
m_{LL} \sim
\left( \begin{array}{lll}
\lambda^4 &  \lambda^2 & \lambda^2\\
\lambda^2 & 1 & 1 \\
\lambda^2 & 1 & 1
\end{array}
\right)m_{\nu_3}
\eeq

with the determinant of the lower 23 sub-block of $m_{LL}$
vanishing to order $O(\lambda^2)$ (as required) 
due to $\bar{\nu}_{3R}$ dominating the contribution
to the 23 block. The reason for $\bar{\nu}_{3R}$ dominance
in this case is that it is lighter than the
next lightest right-handed neutrino $\bar{\nu}_{2R}$
by a factor of $\lambda^2$,
while the Dirac couplings to the second and third lepton
doublets are the same order of magnitude for $\bar{\nu}_{3R}$
and $\bar{\nu}_{2R}$.

(ii) ``Off-Diagonal dominated'' upper block of heavy Majorana matrix. 
This is the kind of model discussed in ref.\cite{Altarelli}.
\beq
 \left( \begin{array}{c} \nu_{1L}\\ e_{1L} \end{array} \right),
 \left( \begin{array}{c} \nu_{2L}\\ e_{2L} \end{array} \right),
 \left( \begin{array}{c} \nu_{3L}\\ e_{3L} \end{array} \right)
=(2,0,0)
\eeq
\beq
\bar{\nu}_{1R},\bar{\nu}_{2R},\bar{\nu}_{3R} = (1,-1,0)
\eeq

The resulting mass matrices are:

\beq
M_{RR} \sim
\left( \begin{array}{lll}
\lambda^2 &  1 & \lambda\\
1 & \lambda^2 & \lambda \\
\lambda & \lambda & 1
\end{array}
\right)M
\eeq

\beq
m_{LR} \sim
\left( \begin{array}{lll}
\lambda^{3} &  \lambda & \lambda^2\\
\lambda & \lambda & 1 \\
\lambda & \lambda & 1
\end{array}
\right)v_2
\eeq

\beq
m_{LL} \sim
\left( \begin{array}{lll}
\lambda^4 &  \lambda^2 & \lambda^2\\
\lambda^2 & 1 & 1 \\
\lambda^2 & 1 & 1
\end{array}
\right)m_{\nu_3}
\eeq

with determinant of the lower 23 sub-block again vanishing to order
$O(\lambda^2)$ (as desired)
due to $\bar{\nu}_{3R}$ dominating the contribution
to the 23 block. The reason for $\bar{\nu}_{3R}$ dominance
in this case is that its Dirac couplings to the second and third lepton
doublets is larger by a factor of $1/\lambda$ compared to those of
$\bar{\nu}_{2R}$, $\bar{\nu}_{1R}$,
while all right-handed neutrinos have roughly equal masses.

(iii)``Democratic'' upper block of heavy Majorana matrix.
\beq
 \left( \begin{array}{c} \nu_{1L}\\ e_{1L} \end{array} \right),
 \left( \begin{array}{c} \nu_{2L}\\ e_{2L} \end{array} \right),
 \left( \begin{array}{c} \nu_{3L}\\ e_{3L} \end{array} \right)
=(3/2,-1/2,-1/2)
\eeq
\beq
{ \bar{\nu}_{1R}},{ \bar{\nu}_{2R}},{ \bar{\nu}_{3R}} = (0,0,1)
\eeq

The resulting mass matrices are:

\beq
M_{RR} \sim
\left( \begin{array}{lll}
1 &  1 & \lambda\\
1 &  1 & \lambda \\
\lambda & \lambda & \lambda^2
\end{array}
\right)M
\eeq

\beq
m_{LR} \sim
\left( \begin{array}{lll}
\lambda^{3/2} &  \lambda^{3/2} & \lambda^{5/2}\\
\lambda^{1/2} & \lambda^{1/2} & \lambda^{1/2} \\
\lambda^{1/2} & \lambda^{1/2} & \lambda^{1/2}
\end{array}
\right)v_2
\eeq

\beq
m_{LL} \sim
\left( \begin{array}{lll}
\lambda^4 &  \lambda^2 & \lambda^2\\
\lambda^2 & 1 & 1 \\
\lambda^2 & 1 & 1
\end{array}
\right)m_{\nu_3}
\eeq

with determinant of the lower 23 sub-block once again vanishing to order
$O(\lambda^2)$ due to $\bar{\nu}_{3R}$ dominating the contribution
to the 23 block. The reason for $\bar{\nu}_{3R}$ dominance in this case
is similar to case (i), namely that it is lighter than the
two other (in this case) degenerate 
right-handed neutrinos by a factor of $\lambda^2$,
while the Dirac couplings to the second and third lepton
doublets are the same for all right-handed neutrinos.

Although we have focussed on the small angle MSW case for
definiteness, similar examples may readily be constructed for
the large angle MSW case. For example case (ii) above may
trivially be extended to the large angle MSW case by taking
\beq
 \left( \begin{array}{c} \nu_{1L}\\ e_{1L} \end{array} \right),
 \left( \begin{array}{c} \nu_{2L}\\ e_{2L} \end{array} \right),
 \left( \begin{array}{c} \nu_{3L}\\ e_{3L} \end{array} \right)
=(1,0,0)
\eeq
\beq
\bar{\nu}_{1R},\bar{\nu}_{2R},\bar{\nu}_{3R} = (1,-1,0).
\eeq

The resulting mass matrices in the large angle case become:

\beq
M_{RR} \sim
\left( \begin{array}{lll}
\lambda^2 &  1 & \lambda\\
1 & \lambda^2 & \lambda \\
\lambda & \lambda & 1
\end{array}
\right)M
\eeq

\beq
m_{LR} \sim
\left( \begin{array}{lll}
\lambda^{2} &  1 & \lambda\\
\lambda & \lambda & 1 \\
\lambda & \lambda & 1
\end{array}
\right)v_2
\eeq

\beq
m_{LL} \sim
\left( \begin{array}{lll}
\lambda^2 &  \lambda & \lambda\\
\lambda & 1 & 1 \\
\lambda & 1 & 1
\end{array}
\right)m_{\nu_3}
\eeq

with determinant of the lower 23 sub-block again vanishing to order
$O(\lambda^2)$ (as desired)
due to $\bar{\nu}_{3R}$ dominating the contribution
to the 23 block. The reason for $\bar{\nu}_{3R}$ dominance
in this case is that its Dirac couplings to the second and third lepton
doublets is again larger by a factor of $1/\lambda$ compared to those of
$\bar{\nu}_{2R}$, $\bar{\nu}_{1R}$,
while all right-handed neutrinos have roughly equal masses, as before.

In conclusion SRHND provides an elegant mechanism for
yielding both large 23 neutrino mixing angles and hierarchical 23
neutrino masses simultaneously by virtue of the approximately
vanishing 23 subdeterminant of $m_{LL}$ in these models.
Such models hence provide a natural explanation
of the atmospheric neutrino data.
U(1) family symmetry is discussed as an 
organising principle which leads to a controlled expansion
in the Wolfenstein parameter $\lambda$, capable of
providing a complete explanation of the quark and lepton spectrum in
general. We give some explicit examples
of U(1) charge assignments in the lepton sector which lead to SRHND
in the 23 block of $m_{LL}$. In these examples, 
the subdeterminant vanishes to
order $\lambda^2$, and gives rise to a non-zero mass ratio
$m_{\nu_2} /m_{\nu_3} \sim \lambda^2$ capable of 
accounting for the solar neutrino data via the large or small angle
MSW effect. In the large angle MSW case
we must rely on numerical factors of order unity to 
slightly enhance $\theta_{12}$ relative to $\theta_{13}$
in order to give a large MSW angle without violating the
CHOOZ constraint.

\end{document}